\documentclass[aps,pra,reprint,superscriptaddress]{revtex4-1}
\usepackage[utf8]{inputenc}

\usepackage{amsmath}
\usepackage{fixmath}

\usepackage{soul}

\usepackage{graphicx}

\begin{document}

\title{Comparison of time profiles for the magnetic transport of cold atoms}

\author{T.~Badr}
\affiliation{CNRS, UMR7538, F-93430, Villetaneuse, France}
\affiliation{Université Paris 13, Sorbonne Paris Cité, Laboratoire de physique des lasers, F-93430, Villetaneuse, France}
\author{D.~Ben Ali}
\affiliation{Université Paris 13, Sorbonne Paris Cité, Laboratoire de physique des lasers, F-93430, Villetaneuse, France}
\affiliation{CNRS, UMR7538, F-93430, Villetaneuse, France}
\author{J.~Seaward}
\affiliation{Université Paris 13, Sorbonne Paris Cité, Laboratoire de physique des lasers, F-93430, Villetaneuse, France}
\affiliation{CNRS, UMR7538, F-93430, Villetaneuse, France}
\author{Y.~Guo}
\affiliation{Université Paris 13, Sorbonne Paris Cité, Laboratoire de physique des lasers, F-93430, Villetaneuse, France}
\affiliation{CNRS, UMR7538, F-93430, Villetaneuse, France}
\author{F.~Wiotte}
\affiliation{Université Paris 13, Sorbonne Paris Cité, Laboratoire de physique des lasers, F-93430, Villetaneuse, France}
\affiliation{CNRS, UMR7538, F-93430, Villetaneuse, France}
\author{R.~Dubessy}
\affiliation{Université Paris 13, Sorbonne Paris Cité, Laboratoire de physique des lasers, F-93430, Villetaneuse, France}
\affiliation{CNRS, UMR7538, F-93430, Villetaneuse, France}
\author{H.~Perrin}
\affiliation{CNRS, UMR7538, F-93430, Villetaneuse, France}
\affiliation{Université Paris 13, Sorbonne Paris Cité, Laboratoire de physique des lasers, F-93430, Villetaneuse, France}
\author{A.~Perrin}
\affiliation{CNRS, UMR7538, F-93430, Villetaneuse, France}
\affiliation{Université Paris 13, Sorbonne Paris Cité, Laboratoire de physique des lasers, F-93430, Villetaneuse, France}
\email[]{aurelien.perrin@univ-paris13.fr}

\date{\today}

\begin{abstract}
We have compared different time profiles for the trajectory of the centre of a quadrupole magnetic trap designed for the transport of cold sodium atoms. Our experimental observations show that a smooth profile characterized by an analytical expression involving the error function minimizes the transport duration while limiting atom losses and heating of the trapped gas: moving the gas over nearly 31~cm requires only about 600~ms. Using numerical calculations of single atom classical trajectories within the trap, we show that this observation can be qualitatively interpreted as a trade-off between two types of losses: finite depth of the confinement and Majorana spin flips.
\end{abstract}


\maketitle

\section{Introduction}

The transport of cold atoms over macroscopic distances is now a well established technique that allows one to spatially isolate two stages in the production of degenerate quantum gases~\cite{Greiner2001}; typically a cold sample is prepared in a magneto-optical trap (MOT) in a first vacuum chamber, and conveyed to a second one with a lower background pressure for a final evaporation stage. This, for instance, gives the opportunity to improve optical and mechanical access where the atoms are manipulated and observed. This can also allow for an increase of the repetition rate of the experiments, with the MOT being loaded while the final part of the experimental sequence is performed.

Various implementations have been explored involving either magnetic or optical fields: a chip magnetic conveyor belt~\cite{Hansel2001,Hansel2001b}, time-varying currents in an assembly of anti-Helmholtz coils~\cite{Greiner2001,Minniberger2014}, optical tweezers~\cite{Gustavson2001,Couvert2008}, a single pair of anti-Helmholtz coils on a translation stage~\cite{Lewandowski2003,Handel2012}, a train of Ioffe-Pritchard traps~\cite{Lahaye2006} or a unidimensionnal optical lattice~\cite{Schrader2001,Schmid2006}. Recently, optimal control has been applied in harmonic~\cite{Torrontegui2011} and anharmonic potentials~\cite{Zhang2015,Zhang2016}. These works allow for the design of fast transport trajectories going far beyond the adiabaticity criterion. In linear traps, the possibility of Majorana spin flips~\cite{Majorana1932} prohibits the existence of adiabatic trajectories which motivates other approaches.

The main objective of this paper is to compare different time profiles for the trajectory of a quadrupole magnetic trap centre and attempt to identify the main factors explaining their performance. In section~\ref{expimp}, we recall the basic principles of magnetically trapping cold atoms in a quadrupole magnetic trap and give details on the experimental design we have used to transport cold atomic gases. In section~\ref{timeprofilescomparison} we investigate different time profiles for the trap centre motion and present our experimental observations. In order to understand our results, we have performed simulations of classical trajectories of the atoms within the moving quadrupole trap. Comparing different time profiles, we propose a qualitative explanation of our experimental results in section~\ref{classicalsimulations}. Finally, section~\ref{conclusion} gives concluding remarks.

\section{Experimental implementation}\label{expimp}
\subsection{Basic principles}\label{basicprinciples}

A straightforward realization of a quadrupole trap can be experimentally obtained with two identical coils in an anti-Helmholtz configuration: in practice this corresponds to two coils separated along their common axis of revolution by a distance comparable to their radii and carrying the same current, $I$, flowing in opposite directions. At the symmetry centre of the assembly, $O$, the produced magnetic field $\mathbf{B}$ vanishes and can be approximated close to this position by a quadrupole field. Assuming $z$ is the axis of revolution of the assembly, it reads
\begin{align}
\mathbf{B}(x,y,z) \simeq \begin{pmatrix} -b' x \\ -b' y \\ 2b' z \end{pmatrix}
\end{align}
where $b'$ is the modulus of the magnetic field gradient in the $x$-$y$ plane. The latter depends on the exact geometry of the coils and is proportional to $I$~\cite{Bergeman1987}. In the following, we keep for simplicity the notation $\mathbf{B}$ to refer to the total magnetic field induced by a given coil configuration.

A set of two pairs of anti-Helmholtz coils with $z$-axis of revolution separated along the $y$-axis by a distance comparable to their radii also produces a quadrupole field  at a position $(0,y_0,0)$ entirely determined by the ratio of the currents flowing in each pair of coils. Close to $(0,y_0,0)$, the resulting magnetic field $\mathbf{B}$ reads
\begin{align}\label{bfield}
\mathbf{B}(x,y,z) \simeq 2b'\begin{pmatrix} -\dfrac{\alpha}{1+\alpha} x \\ -\dfrac{1}{1+\alpha} (y-y_0) \\ z \end{pmatrix}
\end{align}
where $\alpha$ is defined as the ratio of magnetic gradients along the $x$- and $y$-axis and $b'$ is half the modulus of the magnetic field gradient along $z$. The expression in Eq.~(\ref{bfield}) takes into account the fact that $\nabla\cdot \mathbf{B} = 0$ as well as the different symmetries of this particular current distribution.

An atom with magnetic moment $\mathbold{\mu}$ interacts with the magnetic field leading to a coupling $V$. For $^{23}$Na atoms in their ground state $3^2S_{1/2}$ as long as $V$ remains small compared to the hyperfine splittings, one can write
\begin{align}\label{approxV}
V(x,y,z) &= g_Fm_F\mu_B \left| \mathbf{B}(x,y,z) \right| 
\end{align} %% &\simeq 2g_Fm_F\mu_B b' \sqrt{\dfrac{\alpha^2 x^2}{(1+\alpha)^2}+\dfrac{\left(y-y_0\right)^2}{(1+\alpha)^2}+z^2}
where $g_F$ is the Landé factor in the ground state $F$, $\mu_B$ the Bohr magneton and $m_F$ the atomic spin projection onto the local direction of the magnetic field. As soon as $g_Fm_F>0$, $V$ presents a minimum where atoms can be confined. In the following, we neglect gravity since the magnetic gradients $b'$ considered here are sufficiently large and assume $m_F=-1$ since we consider $^{23}$Na atoms trapped in the $|F=1,m_F=-1\rangle$ Zeeman substate. In this case $g_F=-1/2$.

A minimum of three free parameters are necessary to control independently the values of $y_0$, $\alpha$ and $b'$. With two sets of anti-Helmholtz coils, only two currents can be freely and independently tuned. Therefore, it is not possible to move the quadrupole trap along the $(Oy)$ axis while keeping both $\alpha$ and $b'$ constant~\cite{Greiner2001}. Adding a third pair of anti-Helmholtz coils along the $(Oy)$ axis offers an additional degree of freedom which lifts this constraint while keeping the same shape for the magnetic field, $\mathbf{B}$, as in Eq.~(\ref{bfield}).

Overall, the experimental design of a magnetic transport of cold atoms in a quadrupole trap relying on static anti-Helmholtz coils requires a minimum of three independent current supplies. In practice, additional experimental constraints may limit the control over the trap parameters $y_0$, $\alpha$ and $b'$. For instance, our current supplies are not bipolar, so the range of accessible values for $\alpha$ is restricted to values typically larger than $1.5$. We expect that better results can be obtained with bipolar current supplies which would allow to keep $\alpha=1$ throughout the magnetic transport. Moreover, switching smoothly from one set of three pairs of anti-Helmholtz coils to the next requires going through a configuration where only two current supplies out of three deliver a non-zero current. In turn, at these switching positions, only $\alpha$ or $b'$ but not both can be freely set.

\subsection{Experimental design}\label{experimentaldesign}

\begin{figure}[htb]
\centering\includegraphics[width=\linewidth]{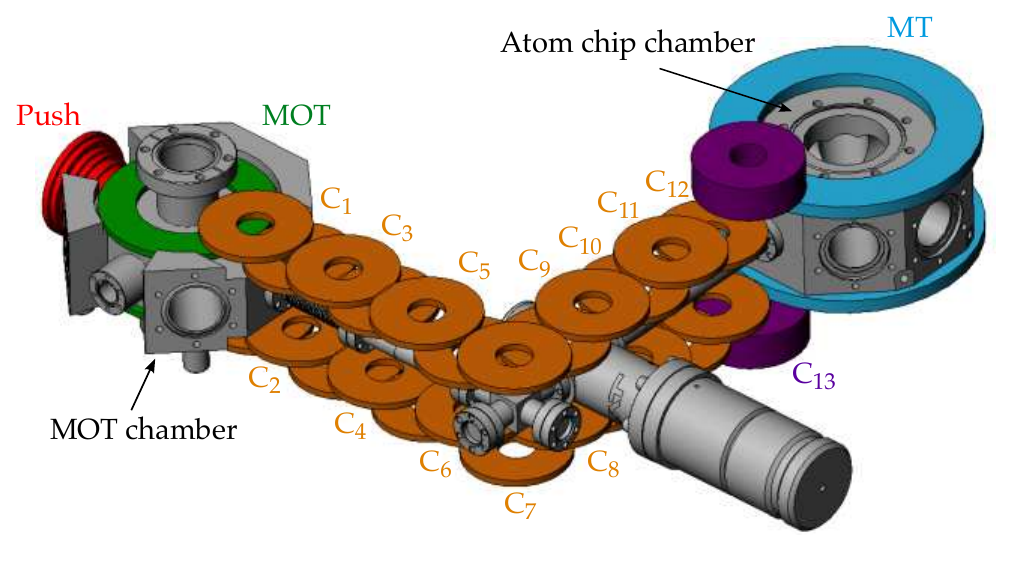}
\caption{\label{figure9}
Overview of the magnetic transport design which connects the MOT chamber to the atom chip chamber. It consists in 15 pairs of anti-Helmholtz coils: MOT, $C_1$ to $C_{13}$ and MT. An additional push coil is used to produce a magnetic gradient which allows us control of the geometry of the quadrupole magnetic trap at the beginning of the transport.}
\end{figure}

The design of our magnetic transport is generally inspired by~\cite{Greiner2001} while its specific implementation is close to~\cite{Minniberger2014}. It relies on 15 pairs of anti-Helmholtz coils and an additional so-called push coil (see Fig.~\ref{figure9}). It allows for transport of atoms along two stages of orthogonal directions and of length $L_1=30.7$~cm and $L_2=34.4$~cm respectively. The atoms are initially confined into a magneto-optical trap (MOT)~\cite{Ali2017} before being transferred into a quadrupole magnetic trap involving only the MOT coils. During the process, $b'$ is ramped up to 65~G/cm and the atoms eventually occupying the $|F=1,m_F=-1\rangle$ Zeeman substate are trapped. At the end of the magnetic transport, the atoms are ready to be transferred onto an atom chip. 

Determining the value of the currents passing through the different coils along the first stage of the magnetic transport requires setting the dependence of $\alpha$ and $b'$ on $y_0$. We have chosen to keep $b'=65$~G/cm constant throughout the magnetic transport. The value of $\alpha$ is equal to one at the beginning and at the end of the first stage, where a single pair of anti-Helmholtz coils is used and the atomic cloud is at rest. As mentioned in the previous section, since we rely on only three non-bipolar current supplies, the value of $\alpha$ is also constrained at each switching position between different sets of three pairs of anti-Helmholtz coils. We explain in Appendix~\ref{annex:currents} how to determine these positions. The value of $\alpha$ is linearly interpolated between these spots apart from the last part of the first stage where only two pairs of anti-Helmholtz coils are then available ($C_6$ and $C_7$) and therefore $\alpha$ evolves freely. This is also the case at the beginning ($C_7$ and $C_8$) and end ($C_{13}$ and MT) of the second stage.

\begin{figure*}[htb]
\centering\includegraphics{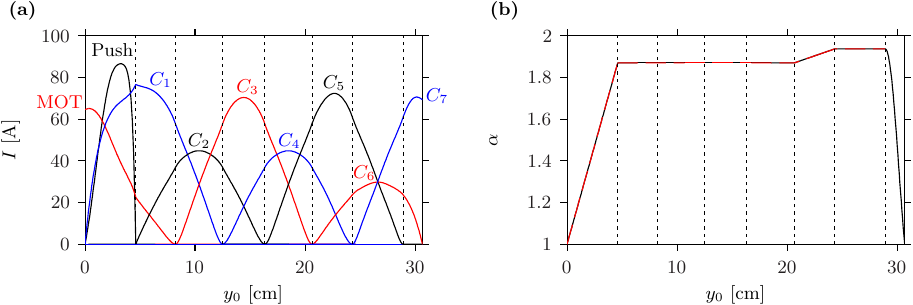}
\caption{\label{figure1}
(a) Currents in the different coils along the first stage of the magnetic transport. Each color corresponds to one of the current supplies. (b) Fitted value of the horizontal trap anisotropy ratio, $\alpha$, along the first stage of the magnetic transport (black line). The red dashed line shows the fit target for $\alpha$ (see Appendix~\ref{annex:currents} for details). In (a) and (b), the black dotted vertical lines indicate the switching positions between two consecutive sets of three pairs of anti-Helmholtz coils. }
\end{figure*}

Relying on the analytical formula of the magnetic field induced by a single current loop~\cite{Bergeman1987} and neglecting the helicity of the coils, it is possible to find an analytical formula for $\left| \mathbf{B}(x,y,z) \right|$ in Eq.~(\ref{approxV}) which takes into account the geometry of each coil and includes their respective number of windings. At each position $y_0$ we then compare this potential to the approximate one arising from Eq.~(\ref{bfield}). Fixing $\alpha$ and $b'$, we fit the current in the different coils so that these two potentials overlap in the best possible way. The results are shown in Fig.~\ref{figure1}. The relative accuracy of the fit on $y_0$, $\alpha$ and $b'$ after this procedure is better than 0.1\%. Additional technical details are given in Appendix~\ref{annex:currents}. The same method is used for the second stage of the magnetic transport.

\section{Time profiles comparison}\label{timeprofilescomparison}

\subsection{Overview}

Controlled displacement of the magnetic trap requires defining the dependence of $y_0$ on the time $t$. In this section we focus on three different trajectories for the trap centre: constant velocity, constant acceleration and an extremely smooth path based on the error function where all time derivatives of the position of the trap are continuous. A constant velocity trajectory leads to the following equations for the position of the trap centre:
\begin{align}\label{cstvel}
y_0(t) &= 0 \ \ \ t\leq 0 \notag\\
y_0(t) &=L_1\frac{t}{\Delta t} \ \ \ 0<t<\Delta t\notag\\
y_0(t) &=L_1 \ \ \ t \geq \Delta t
\end{align}
where $\Delta t$ is the duration of the one-way magnetic transport from the MOT chamber to $C_7$. In this case the acceleration of the trap centre diverges at $t=0$ and $t=\Delta t$ since the velocity is discontinuous. Outside these two points, the effective potential for the atoms in the comoving frame is the same as in the lab frame.

For a constant acceleration trajectory we find:
\begin{align}\label{cstacc}
y_0(t) &= 0 \ \ \ t\leq 0 \notag\\
y_0(t) &=2L_1\left(\frac{t}{\Delta t}\right)^2 \ \ \ 0<t\leq\frac{\Delta t}{2}\notag\\
y_0(t) &=L_1\left[1-2\left(1-\frac{t}{\Delta t}\right)^2\right] \ \ \ \frac{\Delta t}{2}<t<\Delta t\notag\\
y_0(t) &=L_1 \ \ \ t \geq \Delta t.
\end{align}
In this case the velocity of the trap centre is continuous while the acceleration is not. The effective potential for the atoms in the comoving frame becomes tilted so that its gradient along the $(Oy)$ axis is $\mp 2g_Fm_F\mu_Bb'/(1+\alpha)+4mL_1/\Delta t^2$ for $y<y_0(t)$ and $y>y_0(t)$ respectively with $t \in [0,\Delta t/2]$. This is the opposite for $t \in [\Delta t/2,\Delta t]$, which implies an abrupt change in the tilt of the potential in the middle of the trajectory.

Countless other trajectories are conceivable. A few extra examples are given in Appendix~\ref{annex:traj}.  We will focus in the following on a family of trajectories which give good results experimentally. It relies on the error function:
\begin{align}
y_0(t) &= 0 \ \ \ t \leq 0 \notag\\
y_0(t) &=\frac{L_1}{2}\left\{1+\textrm{erf}\left[-\gamma\left(\frac{t}{\Delta t}\right)^{-\delta}+\gamma\left(1-\frac{t}{\Delta t}\right)^{-\delta}\right]\right\}\notag\\
&  \hspace{6.2cm} 0<t<\Delta t \notag\\
y_0(t) &=L_1 \ \ \ t \geq\Delta t
\end{align}
where $\delta>0$ and $\gamma=2^{-3/2-\delta}\sqrt{(\delta+1)(\delta+2)}/\delta$ ensures that the jerk of the trap centre at $\Delta t/2$ vanishes. This allows the acceleration to be close to zero for a large portion of $L_1$. This trajectory is extremely smooth, with continuous derivatives at all orders. The potential in the comoving frame is only tilted at the beginning and the end of the trajectory.

\begin{figure*}[htb]
\centering\includegraphics{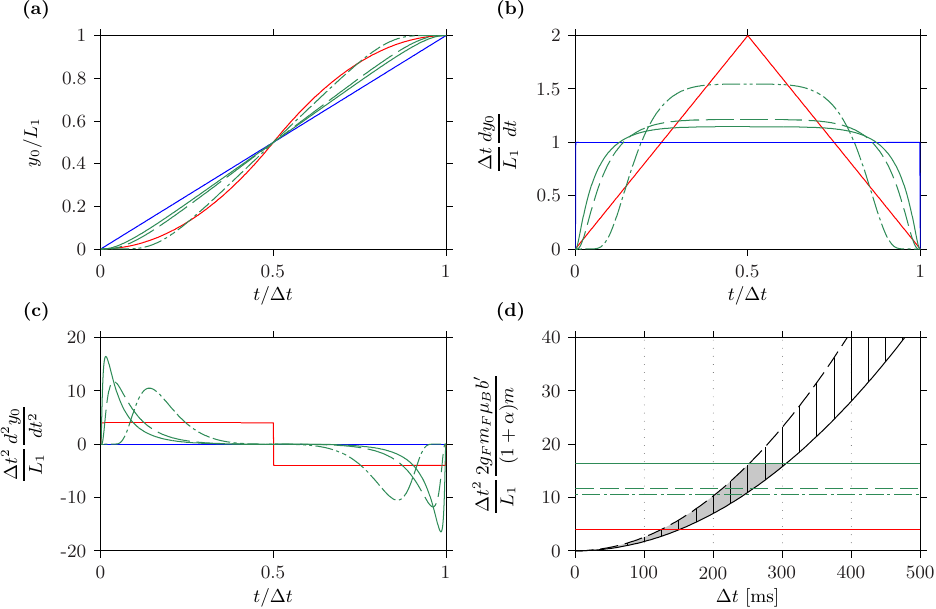}
\caption{\label{figure2}
(a) Behaviour of $y_0$ in units of $L_1$ for different time profiles: constant velocity (blue solid line), constant acceleration (red solid line), error function shape (green lines): $\delta=0.02$ (solid), $\delta=0.1$ (dashed) and $\delta=0.5$ (dashed-dotted). (b) Same as (a) for the velocity of the trap centre expressed in units of $L_1/\Delta t$. (c) Same as (a) for the acceleration of the trap centre expressed in units of $L_1/\Delta t^2$. (d) Absolute value of the acceleration corresponding to the slope of the potential $V$ in units of $L_1/\Delta t^2$: $b'=65$~G/cm and $\alpha=1.937$ (black solid line) and $\alpha=1$ (black dashed line). The vertical black segments indicate the range of slopes covered during a trajectory for some particular values of $\Delta t$. The three horizontal green lines correspond to the maximal absolute value of the acceleration in units of $L_1/\Delta t^2$ of the error function time profiles: $\delta=0.02$ (solid line),  $\delta=0.1$ (dashed line) and  $\delta=0.5$ (dashed-dotted line). The horizontal red line is the same for the constant acceleration profile. As soon as the acceleration corresponding to the slope of the potential $V$ lies below one of the horizontal lines, the atoms become untrapped. As an example, the grey shaded area indicates the range of slopes leading to untrapped atoms for the error function time profile with $\delta=0.02$.}%The grey shaded area indicates the range of slopes covered during the trajectory
\end{figure*}

Figure~\ref{figure2} shows the behaviour of the position (a), velocity (b) and acceleration (c) of the trap centre for different time profiles. Slight changes in the trajectory can result in large modifications of the acceleration. When the absolute value of the latter becomes larger than $2g_Fm_F\mu_Bb'/[(1+\alpha)m]$, the potential $V$ in the comoving frame tilts enough so that the atoms are not trapped anymore. This sets a lower limit on $\Delta t$ above which the atoms remain confined throughout the magnetic transport. This is illustrated in Fig.~\ref{figure2}(d) for the different time profiles considered here. For our typical magnetic gradients and $\alpha\simeq2$ we see that $\Delta t$ must be at least larger than 300~ms for the error function time profile. 

It is important to note that experimentally, the bandwidth of the current supplies is finite which may filter the current output profiles and affect the magnetic transport sequence. We have checked that this effect is negligible in the different situations we have studied.

\subsection{Experimental results}\label{experimentalresults}

\begin{figure*}[htb]
\centering\includegraphics{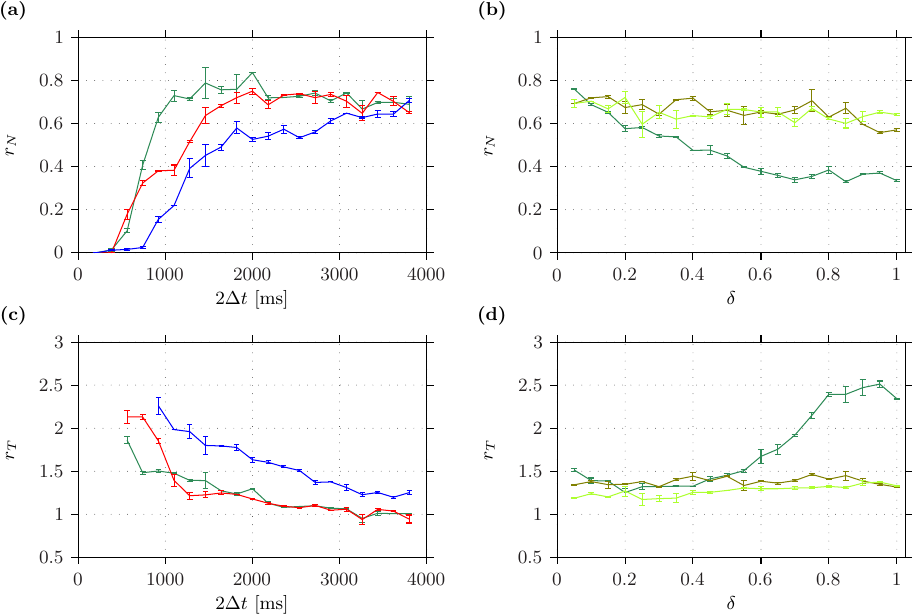}
\caption{\label{figure7}
(a) Ratio $r_N$ of the number of atoms remaining in the magnetic trap after a round trip along the first stage of the magnetic transport for different duration $\Delta t$ and different time profiles: constant velocity (blue solid line), constant acceleration (red solid line) and error function shape with $\delta=0$ (green solid line). (b) Ratio $r_N$ of the number of atoms remaining in the magnetic trap after a round trip along the first stage of the magnetic transport for the error function time profile with different values of $\delta$ and different $\Delta t$: 500~ms (green solid line), 750~ms (olive solid line) and 1000~ms (light green solid line). (c),(d) Same as (a),(b) for the temperature ratio $r_T$.}
\end{figure*}

We have experimentally compared the number of atoms remaining in the magnetic trap after a round trip along the first stage of the magnetic transport for different time profiles and different $\Delta t$. The time profile is just reversed on the way back without any waiting time at the end. In order to account for the losses due to the finite lifetime of the atoms in the trap (about 15~s), we have normalized the results by the number of atoms remaining in the magnetic trap after the same total duration, $2\Delta t$, but without moving; this leads to the ratio $r_N$. Fig.~\ref{figure7}(a) shows that the error function time profile allows us to keep a maximum of about 75\% of the atoms after the round trip and for shorter $\Delta t$ than any other time profile. The worst results are obtained with constant velocity time profile while constant acceleration time profile gives intermediate results.

We have also compared the results of the error function time profile for different $\delta$. The results are shown in Fig.~\ref{figure7}(b). The best results are obtained for the lowest value of $\delta$. Note that $\delta=0$ corresponds to the following trap centre trajectory
\begin{align}\label{erf}
y_0(t) &= 0 \ \ \ t\leq 0 \notag\\
y_0(t) &=\frac{L_1}{2}\left\{1-\textrm{erf}\left[\log\left(\sqrt{\frac{\Delta t-t}{t}}\right)\right]\right\}\notag \quad
   0<t<\Delta t \notag\\
y_0(t) &=L_1 \ \ \ t \geq \Delta t.
\end{align}
Smooth trajectories reaching high values for the acceleration of the trap centre for a short time seem hence to be favoured. This is confirmed by the results of Fig.\ref{figure8}(b), described in Appendix~\ref{annex:traj}. The limit of this strategy comes from the tilt of the potential in the comoving frame: above a certain value of the acceleration the atoms become anti-trapped resulting in large atom losses. 

In order to estimate the heating of the gas due to the magnetic transport, we have measured the temperature $T$ of the cloud with a time-of-flight expansion. Relying on a model described in Appendix~\ref{annex:therm}, we are able to extract $T$ from single shot data. We then normalize the results with the temperature of a gas at rest in the trap for the same total duration $2\Delta t$. This leads to the ratio $r_T$ presented in Fig.~\ref{figure7}(c). Heating is observed for the shortest durations where $r_N$ starts to decrease significantly. The error function time profile gives the best results in particular for the shortest durations. Note that our clouds are not at thermal equilibrium right after the loading of the magnetic trap. Since the collision rate in the trap is low (see Appendix~\ref{annex:col}) the gas barely reaches thermal equilibrium even for the longest transport durations. Because of this, the temperature we extract from our data is strictly speaking an effective temperature and can be different along the horizontal and vertical direction. Nevertheless, our estimation of $r_T$ should still be accurate. Throughout the paper, $r_T$ is estimated from temperature fits along the horizontal direction. 

\section{Classical simulations}\label{classicalsimulations}

\begin{figure*}[htb]
\centering\includegraphics{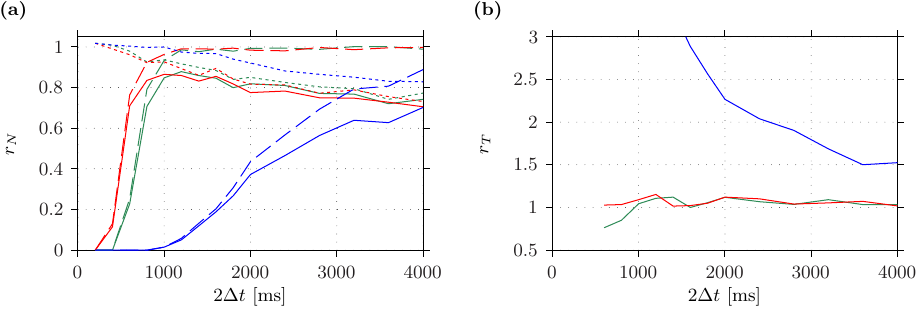}
\caption{\label{figure3}
(a) Ratio $r_N$ of the number of atoms remaining in the magnetic trap after a simulated round trip along the first stage of the magnetic transport for different duration $\Delta t$ and different time profiles: constant velocity (blue solid line), constant acceleration (red solid line) and error function shape with $\delta=0$ (green solid line). The dashed lines show the same ratio but considering only finite trap depth losses. The dotted lines only take into account losses associated with Majorana spin flips. (b)  Temperature ratio $r_T$ estimated from the simulated velocity distribution. The color code is the same than in (a).}
\end{figure*}

Two loss sources can be considered in order to explain our observations: first, losses due to the finite depth of the magnetic trap. Second, Majorana losses due to the fact that the atomic spins cannot adiabatically follow the changes in the magnetic field orientation~\cite{Majorana1932}. In this section we simulate classical trajectories of the atoms in the moving trap in order to compare different loss types for the different time profiles.

Considering each atom as a classical point-like particle, the equation of motion for each atom reads
\begin{align}\label{clsim}
m\frac{d^2\mathbf{r}_a}{dt^2}=-\nabla V(y_0(t),t)
\end{align}
where $\mathbf{r}_a(t)=(x_a(t),y_a(t),z_a(t))$ is the position of an atom at time $t$ and the potential $V$ is fully determined by the three parameters $b'$, $\alpha(y_0)$ and $y_0(t)$ as in Eq.~(\ref{approxV}) with the approximate $\left| \mathbf{B}(x,y,z) \right|$ from Eq.~(\ref{bfield}). In principle, we could have relied on the analytical formula for $V$ taking into account the exact geometry of the different coils but this significantly slows down the numerical calculation and would have required several weeks of computation on a single computer. Moreover we don't expect that this would significantly change our conclusions. In the following, $b'=65$~G/cm and $\alpha(y_0)$ follows the profile depicted in Fig.~\ref{figure1}(b) as in the experiment. The trap centre trajectory, $y_0(t)$, follows a time profile with either a constant velocity, Eq.~(\ref{cstvel}), a constant acceleration, Eq.~(\ref{cstacc}) or an error function with $\delta=0$, Eq.~(\ref{erf}). As a reference, we have also computed atomic trajectories in a static potential $V$ with parameters $b'=65$~G/cm, $\alpha(0)=1$ and $y_0(t)=0$.

With our typical atom number and temperature, the collision rate of the atoms in the trap is smaller than $1$~s$^{-1}$ (see Appendix~\ref{annex:col}). To keep the treatment as simple as possible, in the following we do not take into account interatomic interactions. Despite this choice, we expect our main conclusions to remain qualitatively valid thanks to the low collision rate experienced by the atoms.

In order to estimate the phase-space density of the gas throughout a roundtrip along the first stage of the magnetic transport, we have simulated 1000 different atomic trajectories indexed by a parameter $i$, with initial positions $\mathbf{r}^i_a(t)=(x^i_a(0),y^i_a(0),z^i_a(0))$ randomly picked in order to reproduce a system initially at thermal equilibrium at temperature $T=100~\mu$K in the static trap $V$ with parameters $b'=65$~G/cm, $\alpha(0)=1$ and $y_0(0)=0$ (see Appendix~\ref{annex:num}). This allows us to introduce a length scale $r_0=k_B T/(g_Fm_F\mu_Bb')$, with $k_B$ the Boltzmann constant, and a velocity scale $v_0=\sqrt{k_B T/m}$. The length $r_0$ is related to the size of the atomic cloud at rest while the velocity $v_0$ is simply the rms velocity along each direction of space.

%In order to estimate finite depth losses in the trap associated to each time profile, we have computed for each trajectory $i$ the maximal distance to the trap centre along the $y$ axis $\max\limits_{t} \left|\bar{y}^i_a(t)\right|$ where $\bar{y}^i_a(t)=y^i_a(t)-y_0(t)$. Fig.~\ref{figure3}(a) shows the mean value of this distribution for different transport duration $\Delta t$. The error bars indicate the width of the distribution. We observe that the error function and constant acceleration time profile results converge to the static potential ones as soon as $\Delta t>500$~ms. The constant velocity time profile leads to trajectories which explore a significantly larger space around the trap centre. In our simulations, the potential is idealized and the trap depth is infinite. The atoms are never lost. This is not the case in the experimental realization of $V$, where the radius of the magnetic coils fixes an upper limit to the maximal distance to the trap centre. From the simulations we can then qualitatively expect a larger loss rate due to the finite depth of the trap for the constant velocity time profile than for the other two. The same analysis along the $x$ and $z$ direction gives similar results but with smaller maximal distances to the trap centre.

\begin{figure*}[htb]
\centering\includegraphics{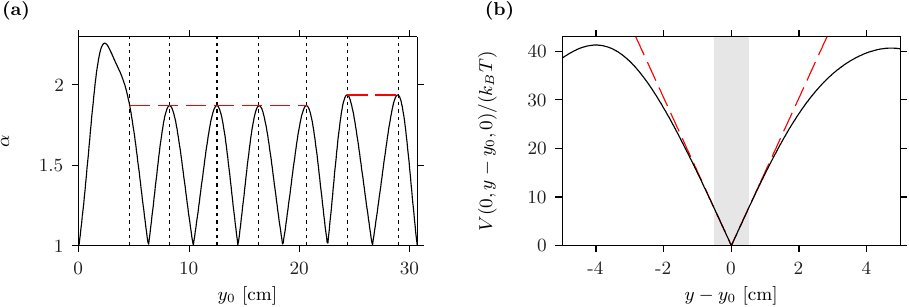}
\caption{\label{figure5}
(a) Dependence of $\alpha$ on the centre position of the magnetic quadrupole trap, $y_0$, when only two consecutive pairs of anti-Helmholtz coils are used. The vertical dotted lines indicate the switching positions from one set of three pairs of anti-Helmholtz coils to the next (see text). The red dashed lines connect the maxima of $\alpha$. (b) Cut in the magnetic quadrupole potential along the $y$ axis around $y_0=L_1/2$. The red dashed line corresponds to $V$ of Eq.~\ref{approxV} taking the approximate $\left| \mathbf{B}(x,y,z) \right|$ from Eq.~\ref{bfield} with $b'=65$~G/cm and $\alpha(y_0)=1.87$. The black line corresponds to the analytical formula of $V$ which takes into account the details of the geometry of each coil. The current flowing in the coils have been adjusted so that the black lines optimally fit the red dashed line within the light grey area (see text).}
\end{figure*}

In order to estimate finite depth losses in the trap associated to each time profile, we have computed for each trajectory $i$ the maximal distance to the trap centre $\bar{r}^i_{\textrm{max}}=\max\limits_{t} \sqrt{x^i_a(t)^2+\bar{y}^i_a(t)^2+z^i_a(t)^2}$ where $\bar{y}^i_a(t)=y^i_a(t)-y_0(t)$. In our simulations, the potential is idealized and the trap depth is infinite. The atoms are never lost. This is not the case in the experimental realization of $V$, where the radius of the magnetic coils fixes an upper limit to the maximal distance to the trap centre (see Fig.~\ref{figure5}(b)). To reproduce this effect, we fix an upper limit $\bar{r}_{\textrm{max}}$ for the maximal distance to the trap centre above which the simulated atom is considered lost.

%Estimating losses associated to Majorana spin flips requires the comparison of two frequencies: the rate associated with changes in the orientation of the magnetic field $\nu^i_{B}(t)=\left\lVert\dfrac{d}{dt}\left\{\dfrac{\mathbf{B}[\mathbf{r}^i_a(t),t]}{ \left\lVert \mathbf{B}[\mathbf{r}^i_a(t),t]\right\rVert} \right\}\right\rVert$ and the Larmor frequency $\nu^i_L(t)=g_Fm_F\mu_B\left\lVert \mathbf{B}[\mathbf{r}^i_a(t),t]\right\rVert/h$. Defining the ratio $r^i_{\textrm{Maj}}=\max\limits_{t} \nu^i_{B}(t)/\nu^i_L(t)$ we can compare the features of the distribution of $r^i_{\textrm{Maj}}$ for the different time profiles. As soon as $r^i_{\textrm{Maj}}$ gets close to 1, the probability of spin flip is high. In Fig.~\ref{figure3}(b) we show that constant velocity trajectories lead on average to smaller values of $r^i_{\textrm{Maj}}$ compared to constant acceleration or error function time profiles. This is explained by the larger extension of the trajectories towards larger magnetic fields and should qualitatively translate into a smaller loss rate associated with Majorana spin flips. Faster magnetic transport also leads to smaller $r^i_{\textrm{Maj}}$. We finally observe that all time profiles lead to larger values of  $r^i_{\textrm{Maj}}$ compared to the static case. This probably explains why the magnetic transport efficiency is never 100\% even for the longest $\Delta t$ in our experimental data.

Estimating losses associated to Majorana spin flips requires the comparison of two frequencies: the rate associated with changes in the orientation of the magnetic field $\nu^i_{B}(t)=\left\lVert\dfrac{d}{dt}\left\{\dfrac{\mathbf{B}[\mathbf{r}^i_a(t),t]}{ \left\lVert \mathbf{B}[\mathbf{r}^i_a(t),t]\right\rVert} \right\}\right\rVert$ and the Larmor frequency $\nu^i_L(t)=g_Fm_F\mu_B\left\lVert \mathbf{B}[\mathbf{r}^i_a(t),t]\right\rVert/h$. We define the ratio $r^i_{\textrm{Maj}}=\max\limits_{t} \nu^i_{B}(t)/\nu^i_L(t)$. As soon as $r^i_{\textrm{Maj}}$ gets close to 1, the probability of spin flip is high. As for the distance to the trap centre, we set an upper limit $r^{\textrm{max}}_{\textrm{Maj}}$ for this ratio. Majorana losses and finite trap depth losses can be treated independently since they correspond in principle to different types of atoms: Majorana spin flips happen mostly with slow atoms spending too much time in the low magnetic field regions while finite trap depth losses arise from hot atoms reaching the upper limit of the trapping potential.

Relying on the limits $\bar{r}_{\textrm{max}}$ and $r^{\textrm{max}}_{\textrm{Maj}}$, we calculate the ratio $r_N$ corresponding to the ratio of simulated atoms remaining in the trap after a round trip along the first stage of the magnetic transport for different duration $\Delta t$ and different time profiles. All the obtained ratios are normalized by the ones obtained from simulations in a static trap for the same duration. The results are shown in Fig.~\ref{figure3}(a) for $\bar{r}_{\textrm{max}}=10 r_0$ and $r^{\textrm{max}}_{\textrm{Maj}}=0.4$. These two values have been set in order to reproduce in the best possible way the experimental results presented in Fig.~\ref{figure7}(a). We see that our simulations qualitatively reproduce the experimental results of Fig.~\ref{figure7}(a) even if they fail to explain the better performance of the error function time profile at short duration $\Delta t$. We see that finite trap depth losses are dominant for short $\Delta t$ whereas losses associated with Majorana spin flips contribute more for long duration of the magnetic transport. We also remark that a constant acceleration trajectory minimizes finite trap depth losses but presents a slightly larger Majorana spin flip rate compared to the error function time profile. Setting an upper limit to the ratio  $r^i_{\textrm{Maj}}$ is probably too crude and a more precise treatment of Majorana losses might potentially lead to a better reproduction of the experimental results for short durations.

%In order to estimate the heating induced by the different time profiles we have calculated the distribution of distances to the trap centre at the end of the magnetic transport $\bar{r}^i_a\left(\Delta t\right)$ where $\bar{r}^i_a(t)=\sqrt{(x^i_a(t))^2+(y^i_a(t)-y_0(t))^2+(z^i_a(t))^2}$. We have also computed the  velocity distribution $\bar{v}^i_a\left(\Delta t\right)$ where $\bar{v}^i_a(t)=\sqrt{\left(\dfrac{dx^i_a}{dt}\right)^2+\left(\dfrac{dy^i_a}{dt}-\dfrac{dy_0}{dt}\right)^2+\left(\dfrac{dz^i_a}{dt}\right)^2}$. The mean value of the distribution for a given $\Delta t$ is displayed in Fig.~\ref{figure3}(c) and (d). The error bars indicate the width of the distribution. Constant velocity time profiles lead to larger mean distances and mean velocities than any other time profiles. The widths of the distributions follow the same behaviour. This can be seen as a signature of larger heating. For the constant acceleration and error function time profiles, the heating should be negligible as soon as $\Delta t>600$~ms. These results agree qualitatively well with our experimental observations.

In order to estimate the heating induced by the different time profiles we have calculated the velocity distribution of the simulated atoms after a round trip. This allows us to estimate the kinetic energy of the atomic distribution that we normalize by the one obtained from the simulations in the static trap for the same duration to obtain the ratio $r_T$. In this analysis, we only consider the simulated atoms that remain in the trap after a round trip taking into account the limits $\bar{r}_{\textrm{max}}$ and $r^{\textrm{max}}_{\textrm{Maj}}$.  The results are shown in Fig.~\ref{figure3}(b). It roughly agrees with our experimental results presented in Fig.~\ref{figure7}(c): the constant velocity time profile leads to significantly larger temperatures compared to the other two trajectories. At short duration of the magnetic transport we do not observe heating in the simulations for the constant acceleration and error function time profiles. This might come from the fact that we do not take into account any collision in the simulation. Even if the collision rate is weak, it tends to redistribute the energy acquired during the acceleration phases among the atoms.

Overall, these simulations tend to indicate that the error function time profile realizes a trade-off between the two sources of losses we have taken into account. This is in qualitative agreement with our experimental results and is probably sufficient to explain why the error function time profile leads to the best magnetic transport efficiency even if our simple classical simulations fail to reproduce this point. A more careful analysis taking into account the collisions and the exact behaviour of Majorana spin flips would be required to fully check this point.

\section{Conclusion}\label{conclusion}

Comparing different time profiles, we have identified an efficient trajectory for the centre of a quadrupole trap designed for the transport of cold sodium $^{23}$Na atoms. It relies on a smooth profile parametrized by the error function. Relying on classical simulations of individual trajectories of the atoms during the transport, we have been able to qualitatively investigate our experimental results: two main loss sources - finite depth of the trap and Majorarana spin flips - limit the efficiency of the magnetic transport for the shortest durations. Constant velocity trajectories tend to minimize the amount of Majorana spin flips while constant acceleration ones optimize the finite trap depth losses. Faster magnetic transport also tends to minimize Majorana losses while finite trap depth becomes a bigger limitation for all time profiles. The error function trajectory corresponds to a trade-off between these two types of losses with a sharp but finite acceleration at the beginning and the end of the transport and an almost constant velocity in between. Overall, we are able to transport on the order of $70\%$ of the atoms over a roundtrip along the first stage of the magnetic transport (nearly 30~cm) in about $2\times600$~ms with limited heating of the gas. We have also tested that the same results could be obtained along the beginning of the second stage of the magnetic transport.

While this work does not answer the question of what is theoretically the optimal trajectory to transport cold atoms in a quadrupole trap over large distances, it gives good hints of the direction where to look for. We hope this will contribute to stimulate theoretical works relying on optimal control to determine the best transport trajectories in linear traps (see~\cite{Sorensen2016,Guery-Odelin2019} and references therein). 

\appendix

\section{Experimental details}

\begin{figure*}[htb]
\centering\includegraphics{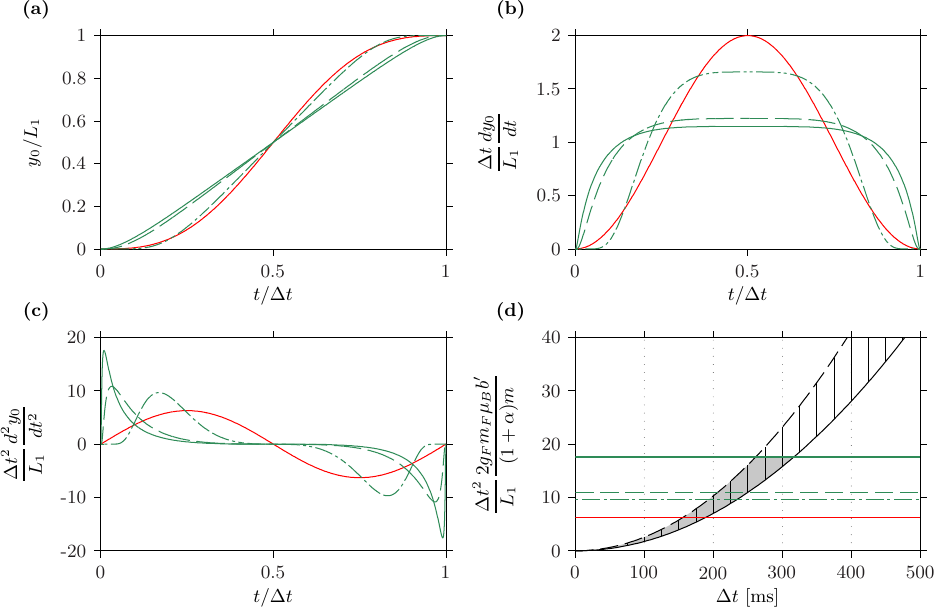}
\caption{\label{figure6}
(a) Behaviour of $y_0$ in units of $L_1$ for different time profiles: sinusoidal acceleration (red solid line), hyperbolic tangent function shape (green lines): $\delta=0.2$ (solid), $\delta=0.3$ (dashed) and $\delta=0.9$ (dashed-dotted). (b) Same as (a) for the velocity of the trap centre in units of $L_1/\Delta t$. (c) Same as (a) for the acceleration of the trap centre expressed in units of $L_1/\Delta t^2$. (d) Absolute value of the acceleration corresponding to the slope of the potential $V$ in units of $L_1/\Delta t^2$: $b'=65$~G/cm and $\alpha=1.937$ (black solid line) and $\alpha=1$ (black dashed line). The vertical black segments indicate the range of slopes covered during a trajectory for some particular values of $\Delta t$. The three horizontal green lines correspond to the maximal absolute value of the acceleration in unit of $L_1/\Delta t^2$ of the hyperbolic tangent function shape time profiles: $\delta=0.2$ (solid line),  $\delta=0.3$ (dashed line) and $\delta=0.9$ (dashed-dotted line). The horizontal red line is the same quantity for the sinusoidal acceleration profile. As soon as the acceleration corresponding to the slope of the potential $V$ lies below one of the horizontal line, the atoms become untrapped. As an example, the grey shaded area indicates the range of slopes leading to untrapped atoms for the hyperbolic tangent function time profile with $\delta=0.2$.}
\end{figure*}

The different coils are made of flat copper wires of rectangular cross section $1\times2.5$~mm$^2$. They are insulated with a thin Kapton\textregistered~layer. The MOT and $C_l$ coils, $l\in[1,12]$, are made of 2 layers of 22 windings with outer and inner diameters of 72~mm and 26.6~mm respectively. Each $C_{13}$ coil is made of 5 $C_l$ coils soldered on top of each other. Each magnetic trap coil (MT) is made of 2 coils soldered on top of each other with 2 layers of 27 windings with an inner diameter of 124~mm. The push coil is conical and made of 16 layers with windings from 2 to 16 and an inner diameter of 38~mm. Except for the latter, all the coils are mounted in a water-cooled aluminium frame.

Three current supplies (one SM 15-100 and two SM 60-100 models from DELTA ELECTRONIKA) are used to deliver the currents in the different coils. An electronic box relying on MOSFETs allows to quickly switch from one pair of coils to another one. The open/close sequences can be read from a rewritable component of the box, or they can be delivered as digital signals by an ADwin-Pro II system with a clock period of $4~\mu$s. The latter also provides the analog signals setting the output current of the different supplies at all time steps.

\begin{figure*}[htb]
\centering\includegraphics{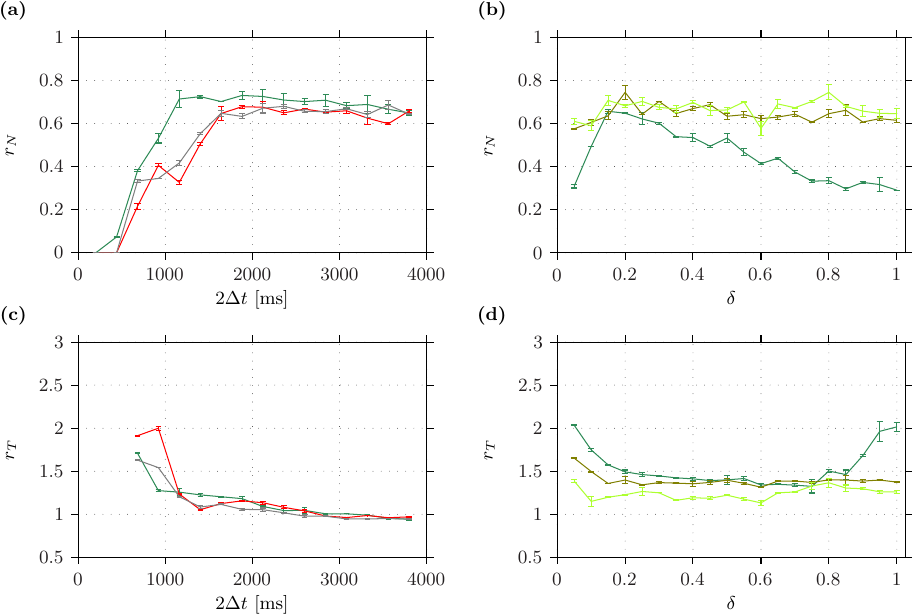}
\caption{\label{figure8}
(a) Ratio $r_N$ of the number of atoms remaining in the magnetic trap after a round trip along the first stage of the magnetic transport for different duration, $\Delta t$, and different time profiles: sinusoidal acceleration (red solid line), constant acceleration (grey solid line) and hyperbolic tangent function shape with $\delta=0.3$ (green solid line). (b) Ratio $r_N$ of the number of atoms remaining in the magnetic trap after a round trip along the first stage of the magnetic transport for the hyperbolic tangent function time profile with different values of $\delta$ and different $\Delta t$: 500~ms (green solid line), 750~ms (olive solid line) and 1000~ms (light green solid line). (c),(d) Same as (a),(b) for the temperature ratio $r_T$.}
\end{figure*}

\section{Determination of the currents}\label{annex:currents}

In order to determine the positions where we switch from one set of three pairs of anti-Helmholtz coils to the next, we first considered a situation where we only use two pairs of anti-Helmholtz coils to move the atoms. Fixing $b'=65$~G/cm, we fit the currents of the two pairs of anti-Helmholtz coils which lie on both sides of $y_0$. This has the advantage of leading to positive current solutions only. If $y_0$ corresponds to the position of the symmetry axis of a given pair, the current in the other pair has to be zero and $\alpha(y_0)=1$. As shown in Fig.~\ref{figure5}(a), between two of these spots, $\alpha$ reaches a maximum: $\alpha=1.87$ for the second to the fifth maximum, and $\alpha=1.937$ for the sixth and seventh. The difference between these two values comes from the fact that the distance between the coils and the magnetic transport axis is slightly larger for the last three pairs of anti-Helmholtz coils of the assembly. These positions of maximal $\alpha$ are the ones we use in our design to switch from one set of three pairs of anti-Helmholtz coils to the next due to the fact that adding a third pair of anti-Helmholtz coils to adjust the shape of the quadrupole trap $V$ can only lead to an increase of $\alpha$ if we restrict ourselves to positive currents. This is actually the opposite with the push coil for which the current direction is set so that positive currents in the push coil lead to lower values of $\alpha$. This is why the first switching position in Fig.~\ref{figure5}(a) is set where $\alpha=1.87$, as in the next ones, and not to the maximum value of $\alpha$ reached with these two pairs of anti-Helmholtz coils.

In order to adjust the shape of the quadrupole trap $V$ throughout the magnetic transport, we fit the analytical formula for $V$ discussed in section~\ref{expimp} to the approximate profile given by Eq.~\ref{approxV} with $b'=65$~G/cm and $\alpha(y_0)$ following the red dashed profile in Fig.~\ref{figure1}(b). More precisely we compute $V(x,y_0,0)$, $V(0,y_0+y,0)$ and $V(0,y_0,z)$ for $x,y,z \in[$-5~mm, 5~mm] with a spatial grid of 101 points along each direction and rely on a least square algorithm to minimize the distance between the potential obtained by the analytical formula and the one deduced from Eq.~\ref{approxV}. The currents in the three supplies are the only free parameters here. For the last few centimetres of the transport, only two current supplies are used and only the profile of $V$ along the $(Oz)$ axis is used. A typical result is illustrated in Fig.~\ref{figure5}(b).

\section{Other examples of trajectories}\label{annex:traj}

We have tested a few additional trajectories in order to complete our experimental observations presented in section~\ref{experimentalresults}. Their respective behaviour is shown in Fig.~\ref{figure6}. The first time profile realizes a sinusoidal profile for the acceleration:
\begin{align}\label{sin}
y_0(t) &= 0 \ \ \ t\leq 0 \notag\\
y_0(t) &=L_1\left[\frac{t}{\Delta t}-\frac{1}{2\pi}\sin\left(2\pi\frac{t}{\Delta t}\right)\right] \ \ \ 0<t<\Delta t\notag\\
y_0(t) &=L_1 \ \ \ t \geq \Delta t.
\end{align}
Such trajectory allows us to check whether the abrupt change in the acceleration in the constant acceleration time profile is critical or not. Fig.~\ref{figure8}(a) actually shows that the answer is negative since it is hard to distinguish the performances of the two time profiles.

We have also tested a time profile very similar to Eq.~\ref{erf} but replacing the error function by a hyperbolic tangent:
\begin{align}
y_0(t) &= 0 \ \ \ t \leq 0 \notag\\
y_0(t) &=\frac{L_1}{2}\left\{1+\textrm{tanh}\left[-\gamma\left(\frac{t}{\Delta t}\right)^{-\delta}+\gamma\left(1-\frac{t}{\Delta t}\right)^{-\delta}\right]\right\}\notag\\
&  \hspace{6.2cm} 0<t<\Delta t \notag\\
y_0(t) &=L_1 \ \ \ t \geq\Delta t.
\end{align}
Such trajectory converges toward a constant velocity time profile when $\delta$ tends to zero. In  Fig.~\ref{figure8}(b), we observe an optimum around $\delta\simeq0.2$ for short values of $\Delta t$. This is in agreement with Fig.~\ref{figure6}(d), where we qualitatively see that for $\delta>0.2$ the duration $\Delta t$ must be larger than 500~ms so that the atoms remain trapped throughout the transport.

\section{Collision rate}\label{annex:col}

The collision rate $\gamma_c$ can be defined as~\cite{dalib1998}
\begin{align}
\gamma_c = \left\langle n \right \rangle \left\langle v_r\right\rangle \sigma
\end{align}
where $\left\langle n \right \rangle$ is the mean density in the trap $V$, $ \left\langle v_r\right\rangle$ the average relative collision velocity and $\sigma=8\pi a^2$ the scattering cross-section with $a$ the scattering length. The scattering length for sodium atoms in the $|F=1,m_F=-1\rangle$ Zeeman substate is $a=2.75$~nm~\cite{Tiesinga1996}. For a cloud at thermal equilibrium in the trap $V$ with $\alpha=1$, one finds
\begin{align}
\left\langle n \right \rangle = \frac{N}{32\pi r_0^3} \hspace{1cm} \left\langle v_r\right\rangle = \frac{4}{\sqrt{\pi}}v_0
\end{align}
where N is the total atom number in the trap.

\begin{figure}[htb]
\centering\includegraphics{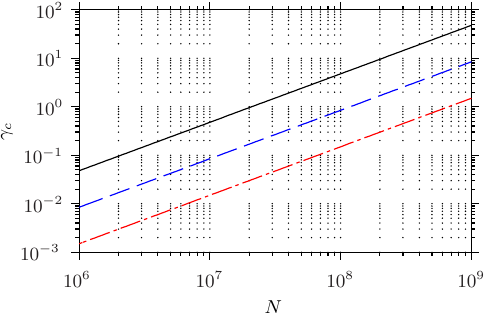}
\caption{\label{figure4}
Collision rate $\gamma_c$ in the quadrupole magnetic trap $V$ with parameters $b'=65$~G/cm and $\alpha=1$ for different atom number $N$ and for a temperature $T=50~\mu$K (black solid line), $T=100~\mu$K (blue dashed line) and $T=200~\mu$K (red dashed-dotted line).}
\end{figure}

We show in Fig.~\ref{figure4} how $\gamma_c$ depends on $N$. For our typical atom number and temperature, $\gamma_c\lesssim1$~s$^{-1}$. This is in good qualitative agreement with our experimental observations of the thermalization time of our gas after loading into the magnetic quadrupole trap which is on the order of a few seconds, where $\sim 3\gamma_c^{-1}$ is expected from numerical simulations~\cite{dalib1998}.

\section{Thermometry}\label{annex:therm}

We estimate the temperature of the trapped gas by relying on a fit of the density profile of the atoms after a time of flight $t_{\rm tof}$. If we assume the atoms to be initially at thermal equilibrium at a temperature $T$, the density in the quadrupole trap $n$ is proportional to $\exp\left[-V/(k_B T)\right]$. More precisely, assuming $\alpha=1$, we have
\begin{align}
 n(x,y,z) =  \frac{N}{4\pi r_0^3}\exp\left[-\frac{\sqrt{x^2+y^2+4z^2}}{r_0}\right].
\end{align}
The velocity of the atoms $\mathbf{v}$ simply follows a Maxwell-Boltzmann distribution
\begin{align}
p(\mathbf{v})=\frac{1}{\left(\sqrt{2\pi}v_0\right)^{3}}\exp\left(-\frac{\mathbf{v}^2}{2v_0^2}\right).
\end{align}
Neglecting interatomic collisions during the expansion of the gas, the density distribution after a time of flight $t_{\rm tof}$ stems from the free expansion of the atoms
\begin{align}\label{denstof}
 n(\mathbf{r};t_{\textrm{tof}}) =  \frac{1}{t_{\textrm{tof}}^3}\int d\mathbf{u} \ n(\mathbf{u}) p\left(\dfrac{\mathbf{r}-\mathbf{u}}{t_{\textrm{tof}}}\right).
\end{align}
While Eq.~\ref{denstof} does not simplify into a simple analytical expression, integrating $n(\mathbf{r};t_{\textrm{tof}})$ along the $y$ and $z$ axis leads to
\begin{widetext}
\begin{align}\label{intdens}
n_{yz}(x;t_{\textrm{tof}}) &=\frac{\beta_t N}{2\sqrt{2\pi}r_0}\exp\left(-\frac{x^2}{2\beta_t^2r_0^2 }\right)+\frac{N}{8r_0}\exp\left(\frac{\beta_t^2}{2}\right)\left[\exp\left(-\frac{x}{r_0}\right)\left(1+\frac{x}{r_0}-\beta_t^2\right)\textrm{erfc}\left(\frac{\beta_t}{\sqrt{2}}-\frac{x}{\sqrt{2}\beta_t r_0}\right)\right. \notag\\
& \hspace{6cm}\left.+\exp\left(\frac{x}{r_0}\right)\left(1-\frac{x}{r_0}-\beta_t^2\right)\textrm{erfc}\left(\frac{\beta_t}{\sqrt{2}}+\frac{x}{\sqrt{2}\beta_t r_0}\right)\right]
\end{align}
\end{widetext}
where $\beta_t=v_0t_{\textrm{tof}}/r_0$ and $\textrm{erfc}$ is the complementary error function. Integrating $n(\mathbf{r};t_{\textrm{tof}})$ along the $x$ and $y$ axis leads to an expression for $n_{xy}$ similar to Eq.~\ref{intdens} but where $x$ has to be replaced by $z$, $r_0$ by $r_0/2$ and $\beta_t$ by $2\beta_t$. For long times of flight such as $\beta_t \gg 1$, $n_{xy}$ and $n_{yz}$ both converge toward a simple Gaussian function with RMS width $\beta_t r_0 = v_0t_{\textrm{tof}}$.

To experimentally measure the temperature of our atomic clouds, we switch off the magnetic trap and let the atoms expand for a few milliseconds. Relying on absorption imaging along the $y$ axis we obtain the integrated density profile $\int dy \ n(\mathbf{r};t_{\textrm{tof}})$. Integrating numerically along either the $x$ and $z$ axis we can then fit the resulting profile with the analytical expression of $n_{xy}$ or $n_{yz}$. This then gives us a measurement of the temperature of the cloud $T$ in a single shot.

\section{Numerical calculations}\label{annex:num}

Single trajectories from Eq.~\ref{clsim} are solved with Matlab\textregistered~relying on the \verb|ode45| solver which is based on the Dormand-Prince method. The relative tolerance for the calculation is set to $10^{-6}$.

\setcounter{section}{1}

\begin{acknowledgments}
We are grateful to D. Guéry-Odelin for helpful comments on the manuscript. We thank A. Kaladjian for the building of a large part of the mechanical pieces we have used for the fabrication of the coils and their mounts. This work has been supported by the ANR Project No.~11-PDOC-021-01 and the Région Île-de-France in the framework of DIM NanoK (Des atomes froids aux nanosciences) project FluoStrong. LPL is a member of DIM SIRTEQ (Science
et Ingénierie en Région Île-de-France pour les Technologies Quantiques).
\end{acknowledgments}

\section*{References}
\bibliography{biblio}

\end{document}